# Electric field control of magnetization dynamics in ZnMnSe/ZnBeSe diluted-magnetic-semiconductor heterostructures

M. K. Kneip[1], D. R. Yakovlev[1,*], M. Bayer[1], T. Slobodskyy[2], G. Schmidt[2], and L. W. Molenkamp[2]

[1]*Experimentelle Physik II, University of Dortmund, 44221 Dortmund, Germany*
[2]*Physikalisches Institut der Universität Würzburg, 97074 Würzburg, Germany*

We show that the magnetization dynamics in diluted magnetic semiconductors can be controlled separately from the static magnetization by means of an electric field. The spin-lattice relaxation (SLR) time of magnetic $Mn^{2+}$ ions was tuned by two orders of magnitude by a gate voltage applied to *n*-type modulation-doped (Zn,Mn)Se/(Zn,Be)Se quantum wells. The effect is based on providing an additional channel for SLR by a two-dimensional electron gas (2DEG). The static magnetization responsible for the giant Zeeman spin splitting of excitons was not influenced by the 2DEG density.

PACS:  75.50.Pp,  76.60.Es,  78.20.Ls,  78.55.Et,  85.75.-d

Diluted magnetic semiconductors (DMS), which combine semiconductor electronic properties with a strong enhancement of spin-dependent phenomena due to the presence of magnetic ions, are well-known materials for testing spintronic concepts [1]. The remarkable optical properties of II-VI DMS materials with $Mn^{2+}$ magnetic ions like (Cd,Mn)Te or (Zn,Mn)Se make them very convenient and popular model systems. In DMS heterostructures the carrier spin manipulation arises from the spin-flip exchange scattering of free carriers on the localized magnetic moments of the magnetic ions. A variety of magneto-optical and magneto-transport effects are caused by the strong exchange interaction of the localized $Mn^{2+}$ magnetic moments with the conduction band electrons (*s-d* interaction) and/or valence band holes (*p-d* interaction). Among them are giant Zeeman splitting of the band states, giant Faraday and Kerr rotation and magnetic polaron formation [2].

It is important for applications based on DMS to control *static* and *dynamic* magnetic properties separately. In II-VI DMS the static and dynamic properties of the Mn spin system depend strongly on the Mn concentration and therefore are strongly correlated with each other [3]. However, the underlying mechanisms are different, which makes possible to decouple them. The paramagnetic $Mn^{2+}$ ions substituting cations in II-VI semiconductors have spin 5/2. Neighbouring and next-neighbouring Mn spins interact antiferromagnetically and form clusters. With increasing Mn content, more and more spins become coupled in the clusters and the typical cluster size increases [4]. The static magnetization is mainly determined by paramagnetic Mn spins not bound to clusters. Contrary to that, the magnetization dynamics is dominated by Mn-Mn interactions in antiferromagnetic clusters [5]. The anisotropic exchange interaction due to the Dzyaloshinski-Moriya mechanism is well established as origin for SLR of Mn ions [2].

The dynamics of SLR in $Zn_{1-x}Mn_xSe$ quantum wells (QW) has been studied in detail by time-resolved photoluminescence [6]. The SLR time $\tau_{SLR}$ varies by five orders of magnitude from $10^{-3}$ down to $10^{-8}$ s with increasing Mn content from $x = 0.004$ up to 0.11. It has been shown that the presence of free electrons in a $Zn_{0.996}Mn_{0.004}Se$ QW reduces $\tau_{SLR}$ by an order of magnitude. This is in line with the reported data for $Cd_{1-x}Mn_xTe$-based QWs, where the SLR has been accelerated by the presence of either electrons [7] or holes [8] provided by modulation doping and tuned by laser illumination. Free carriers, being efficiently coupled with both the Mn spins and the phonon system, provide an additional channel for spin and energy transfer from the Mn spin system into the lattice. That opens a possibility to tune the SLR time by adjusting the carrier density in DMS heterostructures. Note, that the static magnetization in DMS QWs is rather independent of the presence of free carriers [9].

We show in this Letter that the gate voltage control of the two-dimensional electron gas (2DEG) density in (Zn,Mn)Se/(Be,Mg)Se QWs allows to vary the SLR time by more than two orders of magnitude, resulting in electric field control of the magnetization dynamics.

The structure was grown by molecular-bean epitaxy on (001)-oriented *n*-doped GaAs substrates overgrown by n-doped GaAs buffer. II-VI layers were nominally undoped, except a 2 nm thick part of $Zn_{0.94}Be_{0.06}Se$ barrier doped with Iodine donors and separated by a 20 nm thick spacer from $Zn_{0.985}Mn_{0.015}Se$ QW. II-VI heterostructures consists of following layers: 30 nm $Zn_{0.92}Be_{0.08}Se$, 55 nm of $Zn_{0.94}Be_{0.06}Se$ (it includes the modulation doped layer), 2 nm $Zn_{0.985}Mn_{0.015}Se$ quantum well, 15 nm of $Zn_{0.94}Be_{0.06}Se$ and 30 nm $Zn_{0.92}Be_{0.08}Se$. A semitransparent gold contact was deposited on the top, so that a gate voltage can be applied along the structure growth axis to tune the 2DED density in the $Zn_{0.985}Mn_{0.015}Se$ QW. Two samples of that type were fabricated and studied, showing very similar results. The 2DEG density $n_e$ in unbiased structure was about $1.5\times10^{11}$ cm$^{-2}$ and was varied by the applied voltage from about $5\times10^{10}$ up to $3.1\times10^{11}$ cm$^{-2}$. We estimate the electron density from the linewidth of the emission line. This method is reliable when the Fermi energy exceeds the inhomogeneous broadening due to allow fluctuations, which is about 4 meV in the studied samples. It allows us to make direct evaluations for $n_e \geq 1.4\times10^{11}$ cm$^{-2}$, while only extrapolations were possible below this value.

To measure the SLR dynamics an all-optical technique was used. The Mn spin system polarized by an external magnetic field was heated by a pulsed laser and the dynamical shift of the photoluminescence (PL) line was detected (for details see Ref. [6]). The photoexcitation was provided by 7 ns laser pulses of the third harmonic of a Nd:YAG laser at 355 nm with a power of 1.3 mW and a repetition rate of 3 kHz. To detect magnetization dynamics between the pulses a *cw* HeCd laser at 325 nm was used additionally. To minimize the heating of the Mn system the *cw* laser was defocused to reach an excitation density below 0.1 W/cm². PL spectra were detected with a gated charge-coupled-device (CCD) camera with a time resolution of 5 ns combined with a 0.5 m spectrometer. Experiments were performed with samples immersed in superfluid helium at a temperature $T = 1.7$ K. External magnetic fields $B$ up to 7 T were applied parallel to the structure growth axis and to the direction of the collected light (Faraday geometry) by a superconducting split-coil.

Typical PL spectra measured under *cw* excitation at magnetic fields of 0 and 3 T are shown in the inset of Fig. 1. The spectra are given for two different gate voltages. The high structural and optical quality of the samples is approved by the narrow linewidth, not exceeding 5 meV. The low energy shift of the PL lines in external magnetic fields is due to the giant Zeeman splitting of the heavy-hole exciton state. The shift value is equal to the one half of the total Zeeman splitting $\Delta E_Z$. A gate voltage of -0.5 V causes a small increase of the PL linewidth due to an increase of electron density and a small decrease of the Zeeman shift at $B = 3$ T due to weak heating of the Mn system by the electrical current through the structure (see Fig. 3 and discussion below). However, the smallness of this current heating does not influence significantly the measurements of the SLR dynamics.

Optical access to the magnetization $M(B,T_{Mn})$ of the Mn spin system can be obtained by means of the giant Zeeman splitting of excitonic states in external magnetic fields [6]:

$$\Delta E_Z(B,T_{Mn}) = \frac{\alpha-\beta}{\mu_B g_{Mn}} M(B,T_{Mn}) \qquad (1)$$

$$M(B, T_{Mn}) = \mu_B g_{Mn} x N_0 S_{eff}(x) \, \text{B}_{5/2} \left[ \frac{5 g_{Mn} \mu_B B}{2 k_B (T_{Mn} + T_0(x))} \right]. \quad (2)$$

The exchange constants in $Zn_{1-x}Mn_x Se$ for the conduction and valence band are $N_0 \alpha = 0.26$ eV and $N_0 \beta = -1.31$ eV, respectively [10]. $x$ is the Mn mole fraction, $N_0$ is the inverse unit-cell volume and $\text{B}_{5/2}$ is the modified Brillouin function. $g_{Mn} = 2$ is the g-factor of $Mn^{2+}$ ion and $\mu_B$ is the Bohr magneton. $T_{Mn}$ is the Mn spin temperature, which in equilibrium is equal to the lattice temperature $T_{Mn} = T$, but may exceed it significantly under external perturbations, such as laser heating [6, 11]. The effective spin $S_{eff}(x)$ and effective temperature $T_0(x)$ allow a phenomenological description of the antiferromagnetic Mn-Mn exchange interaction (values can be found in Ref. [11]).

It follows from Eqs. (1) and (2) that information about the static magnetization can be received from the PL spectra measured under steady-state conditions, e.g. under *cw* excitation. Such results are given in Fig. 1. As shown exemplary for $U = 0$ and -0.6 V, the Zeeman shift for different voltages is rather similar. It amounts to approximately 20 meV at $B$ = 3 T and saturates at about 25 meV for higher fields. Measurements made for several other voltages show a very similar behavior. This experimental behavior confirms that the static magnetization for the studied samples is independent of the gate voltage and therefore is not sensitive to the 2DEG density. This conclusion is in good agreement with the literature data [9]. The solid line in Fig. 1 is a fit of the experimental data for $U = 0$ V by means of Eqs. (1) and (2).

The magnetization dynamics of the Mn spin system can be analysed from the temporal evolution of the Zeeman shift induced by the laser pulses (could be omitted, as mentioned already a few times: in external magnetic fields) [6]. The Mn spin system is heated up during the laser pulses and is cooled down between the pulses back to lattice temperature. The corresponding changes in the Mn temperature can be followed via the energy shift of the PL maxima $\Delta E_{PL}(t)$ relative to its equilibrium position at a fixed magnetic field. The measurements reported here have been performed at $B = 3$ T. In the studied structures with relatively low Mn concentration $x = 0.015$ the SLR time exceeds significantly the laser pulse duration of 7 ns and the typical lifetime of nonequilibrium phonons of about 1 µs. Therefore, the SLR time can be extracted from the decay of the dynamical response [6]. The respective experimental data measured for three gate voltages are plotted in Fig. 2. For a suitable comparison, the data sets are normalized to the maximum shift $\Delta E_{PL}^{max}$ and plotted on a logarithmic scale. Monoexponential fits are shown by solid lines together with the corresponding values of $\tau_{SLR}$. The SLR time decreases from 140 µs for a gate voltage of $U = 0.6$ V down to 1 µs for $U = -1.5$ V.

A detailed treatment of the SLR time dependence on the voltage applied is depicted in Fig. 3(b). For voltages between $U = 0.7$ and -1.5V the SLR time decreases by more than two orders of magnitude from 160 down to 1 µs. It is therefore evident that the magnetization dynamics is accelerated by the presence of a 2DEG, whose concentration is tuned by the gate voltage. SLR time as long as 530 µs have been reported for an undoped of (Zn,Mn)Se/(Zn,Be)Se QW with a Mn concentration of $x = 0.015$ (sample #6 in Ref. [6]). Therefore, the overall tunability range of the SLR time approaches three orders of magnitude.

The applied voltage induces a current flow through the structure with typical values given in Fig. 3(a) by closed circles. The current may induce heating of the Mn system, the effects of which should be taken into account when treating the static and dynamic magnetization by means of the optical spectroscopy. To evaluate it for the studied structures we have plotted in the same panel the energy position of PL line maximum as a function of gate voltages at $B = 3$ T. One can see that the current heating effect is relatively small. In a wide voltage range from 0.7 to -0.7 V the energy shift is smaller that 2 meV, which is less than 10% of the Zeeman shift at 3 T. Only for U = -1.2 V the energy shift rises to 5 meV. It is also interesting that independent of the voltage polarity the PL line shifts to high energies, i.e. the shift is dominated by the Mn heating and the possible contribution of the quantum-confined Stark shift is very small. For the measurements of the Mn spin dynamics this temporally constant shift is not significant, because only the evolution of the energy shift with time is analysed.

Let us discuss now the mechanism responsible for the acceleration of SLR in the presence of free carriers. In $Zn_{1-x}Mn_xSe$ with low Mn content the SLR process due to direct coupling of the Mn system to the lattice is slow. Typical values have been measured for undoped $Zn_{1-x}Mn_xSe$ in Ref. [6]. In contrast, free carriers (electrons and holes) are strongly coupled with both the magnetic ions and the phonons [see inset of Fig.3 (b)]. The carrier interaction with magnetic ions is based on the fast spin-flip exchange scattering. Therefore the free carriers in doped structures serve as a bypass channel for the slow direct spin-lattice relaxation. The efficiency of this channel is controlled by the carrier density and can be characterized by a relaxation time $\tau_{2DEG}(n_e)$. The experimentally measured SLR time $\tau_{SLR}$ can be described in the following way:

$$\frac{1}{\tau_{SLR}} = \frac{1}{\tau_{SLR}^{Mn-L}} + \frac{1}{\tau_{2DEG}} \qquad (3)$$

where $\tau_{SLR}^{Mn-L}$ is the time characteristic for the process provided by direct interaction on Mn ions with the phonon system. Obviously $\tau_{SLR}^{Mn-L} = \tau_{SLR}$ in undoped samples. The $\tau_{SLR}^{Mn-L}$ value can be obtained from measurements on nominally undoped samples and $\tau_{2DEG}$ can be then calculated using a formalism developed in Ref. [12]. This opens the possibility to design structures with predefined SLR times.

In this Letter the ability to control the magnetization dynamics of the Mn spin system by an external electric field was demonstrated. The 2DEG acts as an efficient bypass channel for the energy transfer from the Mn-ion system to the lattice. The gate voltage control of the 2DEG density allows us to tune the spin-lattice relaxation rate in (Zn,Mn)Se/(Be,Mg)Se heterostructures by more than two orders of magnitude. This method can be combined with the recently reported possibilities to address magnetization dynamics by special profiles of Mn distribution in either magnetic digital alloys [3] or in heteromagnetic heterostructures [13].

**Acknowledgements** This work was supported by the BMBF program "nanoquit" and by the DFG via Sonderforschungsbereich 410.

**Figure Captions**

Fig. 1  Giant Zeeman shift of photoluminescence line ($\sigma^+$ polarized) for two different gate voltages ($U = 0$ and -0.6 V). The line shifts to lower energies. PL is excited by a *cw* HeCd laser with a power density of 10 mW/cm². Solid line shows a fit along Eq. (2) with $x = 0.015$, $S_{eff} = 2.5$ and $T_0 = 1$ K. PL spectra at different magnetic fields and gate voltages are given in the insert.

Fig. 2  Temporal evolution of PL line shift corresponding to the cooling of the Mn spin system heated by pulsed laser excitation. The SLR times were evaluated from mono-exponential fits given by solid lines.

Fig. 3  (a) Gate voltage dependences of the PL line maxima energy (open circles) and current (closed circles). (b) SLR time dependence on gate voltage. In insert direct and indirect (via 2DEG) channels for spin and energy transfer from Mn system into lattice are shown schematically.

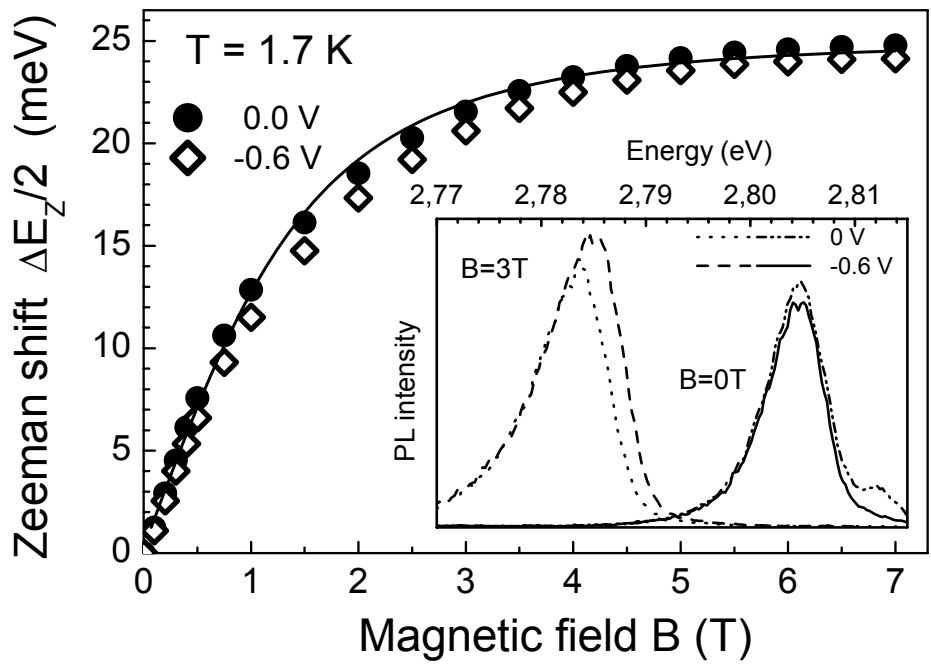

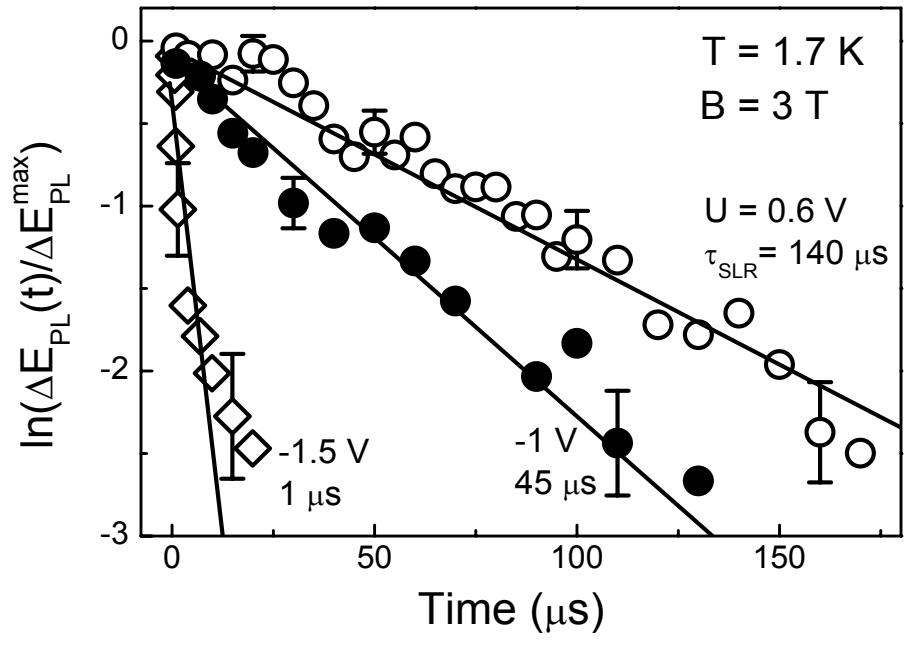

FIG.

(a) B = 3 T, T = 1.7 K

(b) sample 1, sample 2

1 μs

2DEG

$\tau_{2DEG}$

$\tau_{SLR}^{Mn-L}$ Mn spin system

Lattice